\documentclass{article}
\usepackage{spconf,amsmath,graphicx}
\usepackage{booktabs}
\usepackage{algorithm,algorithmicx,algpseudocode}
\usepackage{adjustbox}
\usepackage{subfig}
\usepackage{makecell, tabularx, multirow}
\usepackage{xcolor}
\usepackage{hyperref}
\usepackage{amsfonts}
\usepackage{makecell} 

\title{An Initial Investigation of Neural Replay Simulator for Over-the-Air Adversarial Perturbations to Automatic Speaker Verification}

\name{Jiaqi Li$^{1}$, Li Wang$^{1}$, Liumeng Xue$^{1}$, Lei Wang$^{2}$, Zhizheng Wu$^{1}$ \thanks{This work is supported by NSFC 62376237}}
\address{$^1$School of Data Science, Shenzhen Research Institute of Big Data, \\The Chinese University of Hong Kong, Shenzhen (CUHK-Shenzhen), China \\ $^2$Independent researcher, Singapore}
\begin{document}
\ninept
\maketitle
\begin{abstract}

Deep Learning has advanced Automatic Speaker Verification (ASV) in the past few years. Although it is known that deep learning-based ASV systems are vulnerable to adversarial examples in digital access, there are few studies on adversarial attacks in the context of physical access, where a replay process (i.e., over the air) is involved. An over-the-air attack involves a loudspeaker, a microphone, and a replaying environment that impacts the movement of the sound wave. Our initial experiment confirms that the replay process impacts the effectiveness of the over-the-air attack performance. This study performs an initial investigation towards utilizing a neural replay simulator to improve over-the-air adversarial attack robustness. This is achieved by using a neural waveform synthesizer to simulate the replay process when estimating the adversarial perturbations. Experiments conducted on the ASVspoof2019 dataset confirm that the neural replay simulator can considerably increase the success rates of over-the-air adversarial attacks. This raises the concern for adversarial attacks on speaker verification in physical access applications.


\end{abstract}
\begin{keywords}
Speaker verification, adversarial attack, replay attack, over-the-air attack, replay simulation
\end{keywords}

\section{Introduction}
\vspace{-3mm}
Automatic Speaker Verification (ASV) is to verify a speaker's identity using a presented utterance against an enrolled voice~\cite{wu2015spoofing}. Recently, deep learning-based ASV systems become the mainstream solution~\cite{asv_review_2022}.
However, it is known that deep neural networks are vulnerable to adversarial attacks \cite{survey_adversarial,wang2023advsv}.
These attacks utilize adversarial examples -- instances that have small perturbations maliciously added to a regular speech sample -- to make ASV systems classify incorrectly \cite{adversarial_sv}. Recently, adversarial attacks against ASV systems and their countermeasures have received a growing interest in light of the potential security threats they may cause \cite{tencent_investigating_2020,ntu_adversarial_countermeasure}.

\begin{figure}[htbp]
    \centering
    \includegraphics[width=0.42\textwidth]{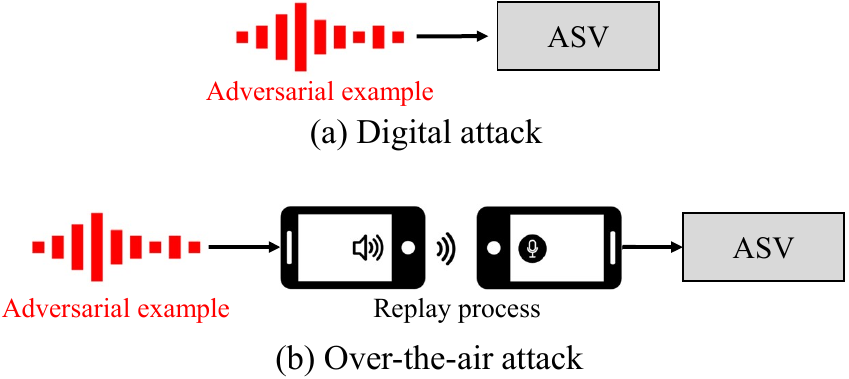}
    \caption{A comparison of digital and over-the-air adversarial attacks to ASV systems.}
    \vspace{-3mm}
    \label{fig:comp}
\end{figure}

There are two types of adversarial attacks, as illustrated in Fig.~\ref{fig:comp} -- \textit{digital} and \textit{over-the-air attacks}. A digital attack is to send a digital copy of an adversarial example to an ASV system, while an over-the-air (OTA) adversarial attack involves replaying an adversarial sample to attack an ASV system. An OTA attack involves a loudspeaker, a microphone, and a replaying environment (e.g., room reverberation and acoustic noise).  There are a number of studies on digital adversarial attacks, which have been found to threaten various ASV architectures~\cite{adversarial_asv_review_interspeech2020,fakeblob}, and the transferability of such attacks from one system to another has been demonstrated \cite{adversarial_asv_review_interspeech2020,fakeblob,adv_training_interspeech2019}. Countermeasures against such attacks have also been investigated, including adversarial training~\cite{ntu_adversarial_countermeasure,adv_training_interspeech2019}, and detection networks~\cite{tencent_investigating_2020,adv_defense_network_icassp2021}.


\begin{figure*}[h!]
    \centering
    \includegraphics[width=0.65\textwidth]{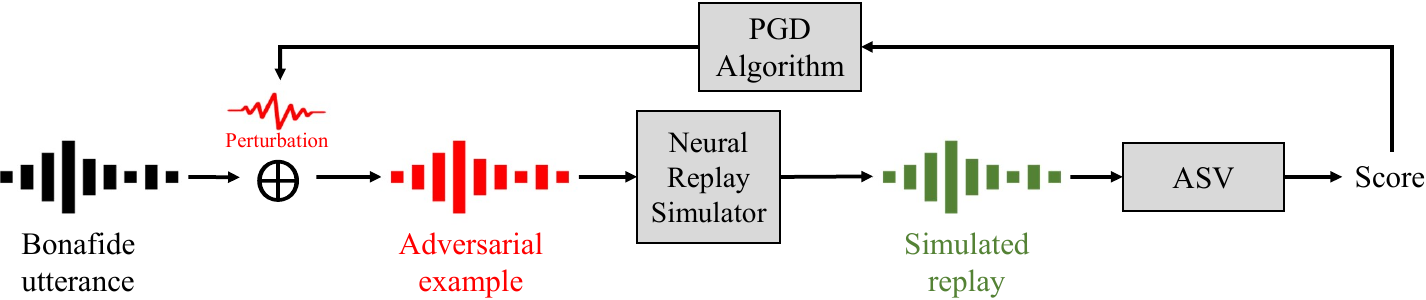}
    \caption{The pipeline to synthesize robust over-the-air adversarial examples utilizing a neural replay simulator. 
    The PGD algorithm synthesizes adversarial perturbation by attacking an ASV model with a pre-trained neural replay simulator. The perturbation is added to a bonafide utterance to become an adversarial example that is expected to remain effective after replay. }
    \label{fig:pipeline}
\end{figure*}

There are few studies on over-the-air adversarial attacks. Some \cite{fakeblob,practical_physical_adv_asr} have observed degraded attack effectiveness of adversarial examples because of the distortion caused by the replay process. To increase the attack effectiveness, a few studies proposed to incorporate a replay simulator into the generation process of perturbation to make the adversarial samples remain effective after replay~\cite{practical_physical_adv_asr,thu_practical_asv_adv_2021}. The main idea behind those studies is to employ a Room Impulse Response (RIR) estimation function to simulate sound reverberation in a room~\cite{practical_physical_adv_asr}. However, those studies mainly focus on the simulation of room reverberation~\cite{pretrain_replay_simualtion_sensors}. It also lacks generalizability for other room sizes and makes it harder to reproduce existing results, since it requires a re-measurement of the RIR for each room environment~\cite{thu_practical_asv_adv_2021,practical_physical_adv_asr}.

To address the limitations of existing approaches, this study proposes a \textit{neural replay simulator}, which focuses on the impact of devices (i.e., the loudspeaker and microphone during the replay process). The neural replay simulator is built using a neural network of U-Net structure by learning from paired clean recording and its recording replayed version. In comparison to the RIR-based approaches, the neural replay simulator is data-driven, and is expected to have a better generalization.  \textbf{\textit{To the best of our knowledge, this is the first attempt to simulate the audio replay process with a machine-learning model.}} It is also the first work to investigate over-the-air adversarial example effectiveness using a neural replay simulator.

\vspace{-3mm}
\section{Over-the-Air Adversarial Perturbations}
\label{sec:pipe_overview}
\label{sec:adv_attack}
\label{sec:our_pipeline}
\vspace{-3mm}

The over-the-air adversarial attack framework with a neural replay simulator is presented in Fig.~\ref{fig:pipeline}. In the framework, there are three major components, namely a neural replay simulator, an automatic speaker verification (ASV) system, and a perturbation generation algorithm. In this section, we will describe each component. We note that the perturbation can be generated by applying the algorithm on both an OTA attack and a digital attack using an ensemble method. This will be discussed in Section~\ref{sec:joint_framework}.

\subsection{Automatic Speaker Verification System}
\vspace{-1mm}

\newcommand{\norm}[1]{\left\lVert #1 \right\rVert}
An ASV system is to decide whether an enrolled speech sample $x_{enroll}$ and a speech sample under test $x$ are uttered by the same speaker. Usually, deep learning-based ASV systems first convert the speech samples $x_{enroll}$ and  $x$ to $n$-dimensional embeddings using a deep network $f$. After that, a cosine similarity score is calculated between $f(x)$ and $f(x_{enroll})$, as $\operatorname{Score}(x, x_{enroll}) = \frac{f(x) \cdot f(x_{enroll})}{\norm{f(x)}_2 \norm{f(x_{enroll})}_2}$.
%
The decision is formed by comparing the score with a fixed verification threshold $\tau$. If Score$(x, x_{enroll}) \geq \tau$, it will output a positive result, indicating that $x$ and $x_{enroll}$ are uttered by the same speaker. Otherwise, a negative result will be the output.
\vspace{-3.5mm}
\begin{figure}[h!]
    \centering
    \includegraphics[width=0.21\textwidth]{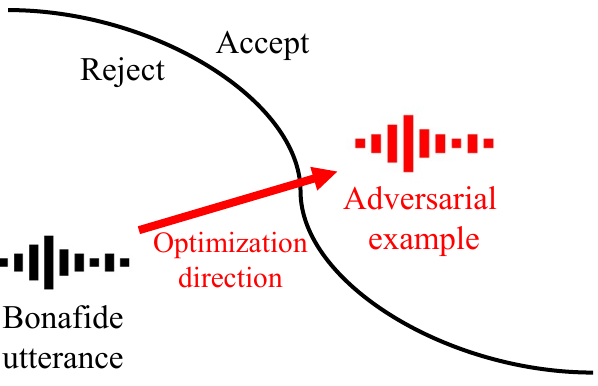}
    \caption{An illustration of an adversarial example cross the decision boundary to manipulate a decision result of an ASV system.}
    \label{fig:decision_boundary}
\end{figure}

The threshold $\tau$ reflects the decision boundary formed by neural networks. It is known that adversarial perturbations can move a negative sample to cross the decision boundary to become a positive sample \cite{advsarial_decision_boundary_2018} as illustrated in Fig.~\ref{fig:decision_boundary}.

\vspace{-2mm}
\subsection{Adversarial Perturbation}
\label{sec:adversarial_perturbation}
\vspace{-1mm}
This study focuses on \textit{targeted} attacks against ASV systems that make an ASV system classify an impostor sample as genuine.
These attacks utilize an adversarial example $x_i$ with some perturbations added to an arbitrary utterance $x_0$.
The adversarial example $x_i$ is expected to deceive the model into classifying that $x_i$ and $x_{enroll}$ are uttered by the same speaker~\cite{imperceptible_asr_simulation_2019}.
The Projected Gradient Descent (PGD)~\cite{pgd} algorithm is employed to synthesize this adversarial example. The algorithm iteratively adds adversarial perturbation to $x_0$ in the increasing direction of an adversarial loss $L_{adv}$ to form a final adversarial example $x_i$,
\begin{equation}
    L_{adv}(x, x_{enroll}, \tau) = \operatorname{Score}(x, x_{enroll}) - \tau
\end{equation}
\begin{equation} \label{equ:fgsm}
    x_{i} \leftarrow x_{i-1} + \epsilon * \operatorname{sign}(\nabla_x L_{adv}(x_{i-1}, x_{enroll}, \tau))
\end{equation}
where $x_i$ denotes the adversarial example at the $i$-th iteration, $x_0$ is a bonafide utterance, $\epsilon$ is the step size, $\nabla_x$ takes the adversarial loss's derivative with respect to the model input $x_i$ and the sign function $sign$. For a fair comparison, this study iterates the PGD algorithm until the first successful attack. In other words, all synthesized $x_i$ achieves 100\% digital attack success rates. 


\vspace{-5mm}
\subsection{Neural Replay Simulator (NRS)} 
\label{sec:components}
\begin{figure}[h!]
    \centering
    \includegraphics[width=0.44\textwidth]{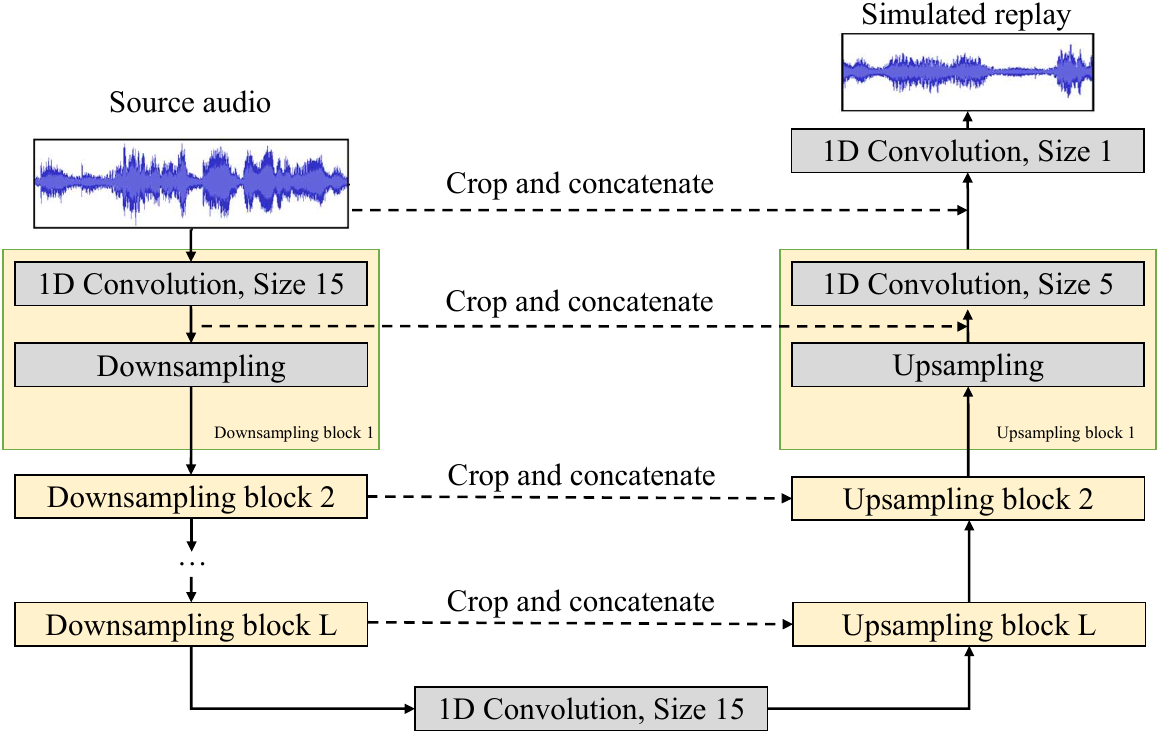}
    \caption{Model structure of the neural replay simulator.}
    \label{fig:waveunet}
\end{figure}

As illustrated in Fig.~\ref{fig:comp}, the replay process is the core module during an over-the-air attack. To increase the success rate of an over-the-air attack, it is important to build a replay simulator. In this study, we utilize Wave-U-Net~\cite{waveunet2018} to build a neural replay simulator. Its architecture is presented in Fig.~\ref{fig:waveunet}. It takes an audio recording as input, and predicts its replayed version. The neural replay simulator is pre-trained with parallel data (i.e., pairs of cleaning recordings and their corresponding replayed version), and is then frozen in the perturbation generation process.

The neural replay simulator is to build a transformation function $\mathcal{F}(\cdot)$ that can transfer the original recording to its replayed version. During training, given a clean recording paired with its replayed version, the Wave-U-Net takes the clean recording as input, and uses its replayed version as the reference target to calculate a loss as its training objective.

The design of the loss function for the neural replay simulator will have a significant impact on the simulation performance. Since the neural replay simulator is closely related to neural vocoders~\cite{bigvgan2023,hifigan2020,melgan2019}, we investigate the loss functions that have proven to work well for neural vocoders as follows:
\begin{itemize}
    \item \textbf{Mel L1}: It is a spectral loss and we use the L1 loss on 80-dimensional Mel-spectrogram extracted from the waveform.
    \item \textbf{Wav L1 and Wav L2}: They are waveform losses following previous researches~\cite{l1_loss_source_sep, waveunet2018}. \textbf{Wav L1} and \textbf{Wav L2} losses employ L1 and L2 loss functions, respectively. 
    \item \textbf{GAN loss}: It is a loss formed by a Multi-Period Discriminator (MPD) and a Multi-Scale STFT (MSSTFT) discriminator same as that in~\cite{gu2023multi}. The implementation of discriminators are from~\cite{zhang2023amphion}. The generator loss is a L1 Mel-Spectrogram loss($L_{\operatorname{Mel}}$) following HiFi-GAN~\cite{hifigan2020}.  The complete training loss is: $L_{\operatorname{GAN}} = \lambda_1 L_{\operatorname{Mel}} + \lambda_2 L_{\operatorname{{MPD}}} + \lambda_3 L_{\operatorname{{MSSTFT}}},$
where $\lambda_1$, $\lambda_2$, $\lambda_3$ are the corresponding weights.
    \item \textbf{ASV loss}: This is a novel task-specific ASV loss to minimize the speaker identity difference between replay simulator output $\hat{y}$ and the ground-truth replay $y$. It is defined as,
    \vspace{-0.5mm}
    \begin{equation}
    L_{\operatorname{ASV}} = \lvert \operatorname{Score}(y, x_{enroll}) - \operatorname{Score}(\hat{y},  x_{enroll}) \rvert ,
    \end{equation}
    where $y$ is a ground-truth replayed audio, $\hat{y}$ is an output from NRS. Our initial experiments show that the ASV loss can be beneficial as an auxiliary loss, but does not yield good results when used alone. Therefore, we explore the loss in a combination form of Mel L1 + ASV loss.
\end{itemize}

Apart from the above loss configurations, we also explore a combined loss of the spectral and time domain (e.g. Mel L1 + Wav L1). When losses are combined, we empirically set each loss's weight $\lambda$ 
and ensure that each loss's contribution to the combined loss is nontrivial. 
For the ASV loss, we use the ECAPATDNN system (to be mentioned in Section \ref{sec:dataset}) as ASV, with $x_{enroll}$ randomly drawn from 100 bonafide audios from the ASVspoof2019~\cite{asvspoof2019} dataset.

When the neural replay simulator is integrated into the adversarial perturbation generation process, the adversarial sample update formula presented in Eq.~\ref{equ:fgsm} can be reformulated as:
\begin{equation} 
    x_{i} \leftarrow x_{i-1} + \epsilon * \operatorname{sign}(\nabla_x L_{adv}(F(x_{i-1}), x_{enroll}, \tau))
\end{equation}

\subsection{Digital and Over-the-Air Joint Attack Framework}
\label{sec:joint_framework}
\vspace{-3mm}
\begin{algorithm}[htbp]
  \caption{Ensemble PGD Attack}
  \label{alg:ensemble_pgd}
  \begin{algorithmic}[1]
    \Require{Surrogate models $\{M_1, M_2, \ldots, M_n\}$, enrollment sample $x_{enroll}$, bonafide utterance $x_0$}
    \Ensure{Adversarial example $x_{adv}$}
    \Function{EnsemblePGD}{$x$}
      \State Initialize $x_{adv} \gets x_0$
      \Repeat
        \For{$i \gets 1$ \textbf{to} $n$}
          \State Compute adversarial sample of \\ \qquad \qquad \qquad $M_i$: $x_{adv} \gets PGD(x_{enroll},x_{adv}) $
        \EndFor
      \Until{$x_{adv}$ successfully attacks all surrogate models}
      \State \Return $x_{adv}$
    \EndFunction
  \end{algorithmic}
\end{algorithm}
\vspace{-5mm}
\begin{figure}[htbp]
    \centering
    \includegraphics[width=0.4\textwidth]{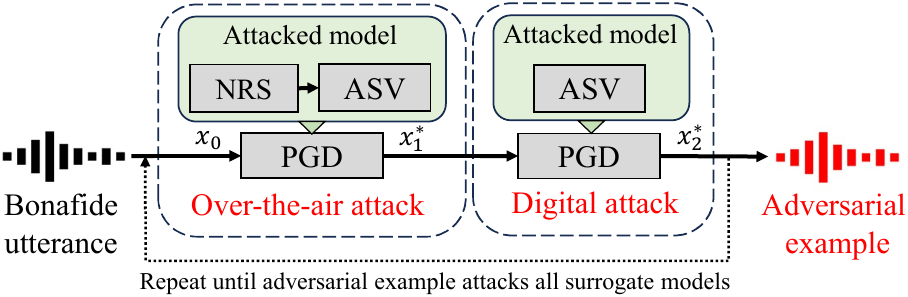}
    \caption{A digital and over-the-air joint attack framework.}
    \label{fig:joint_pipeline}
\end{figure}
\vspace{-1mm}

Inspired by the ensemble-based attack strategy~\cite{ensemble_review2020,ensemble}, we utilize a cascade ensemble approach to perform digital and over-the-air joint attacks in the same framework, expecting a higher success rate for OTA attacks. 
The ensemble PGD algorithm is presented in Algorithm \ref{alg:ensemble_pgd}~\cite{ensemble}, and the framework is illustrated in Fig.~\ref{fig:joint_pipeline}. In our joint attack framework, we set $n=2$ in Algorithm \ref{alg:ensemble_pgd}; $M_1$ be a combined NRS-ASV model, which represents an OTA attack; $M_2$ be the ASV system, which represents a digital attack. In each ensemble PGD step, an OTA attack is performed by attacking the ASV via a replay simulator. This is to synthesize $x_1^{*}$ by adding perturbations to the bonafide input $x_0$. After that, only the ASV model is attacked to synthesize $x_2^{*}$ by perturbing the input $x_1^{*}$. The ensemble PGD algorithm is repeated until the adversarial sample can successfully attack all surrogate models.
The $PGD$ algorithm follows the same configurations as mentioned in section \ref{sec:adversarial_perturbation}.

\vspace{-1.5mm}
\section{Experiments} \label{sec:exp}
\vspace{-3mm}

We conduct experiments for both white-box and transfer attacks. For the white-box attack, it is assumed that the attacker knows everything about the target ASV system, and a victim ASV model is attacked by adversarial examples synthesized with the model itself as the surrogate. In a transfer attack, the attacker has no knowledge about the target ASV system but can access other ASV systems, and the adversarial sample is synthesized from other surrogate ASVs.
\vspace{-2mm}
\subsection{Experimental setup}
\vspace{-2mm}
\label{sec:dataset}
\textbf{Dataset}~ The data we use in this study is derived from the bonafide utterances of the ASVspoof2019~\cite{asvspoof2019} dataset, including a replay set and an attack set. The replay set is mainly used to train the neural replay simulators, while the attack set is used for attack performance evaluation. We note that the data from the two sets are disjoint. 
\begin{itemize}
    \item \textbf{Replay set}: There are 40,000 digital adversarial examples paired with their replayed counterparts. The digital adversarial examples are derived from 2,500 bonafide utterances. We split the dataset into 39,000 and 1,000 examples for NRS training and validation, respectively.
    \item \textbf{Attack set}: 2,500 bonafide utterances from the ASVspoof2019 dataset are used as adversarial attack testing samples. The Attack set is used to generate adversarial examples, which are then recorded in a designed over-the-air attack scenario. During the OTA recording, audio is played by a mid-range Android phone and recorded by a high-end iOS phone at a 0.5-meter distance. The recording is conducted in a soundproof studio to avoid other interfering noises.
\end{itemize}

\noindent \textbf{ASV Systems}~
Four state-of-the-art ASV systems are used. They are all trained with an open-source system, VoxCeleb Trainer~\cite{vox_trainer}. The VoxCeleb2\cite{voxceleb2_2018} dataset is used for training, and VoxCeleb1\cite{voxceleb1} is used for testing. The systems are detailed as follows,
\begin{itemize}
    \item \textbf{XVec:} This is an X-Vector system following the implementation in~\cite{xvector}. The EER is 2.08\%.
    \item \textbf{RawNet:} A RawNet3~\cite{rawnet} system. The EER is 1.06\%.
    \item \textbf{TDNN:} An ECAPATDNN~\cite{ecapa-tdnn} system. The EER is 1.26\%.
    \item \textbf{ResNet:} A ResNetSE34V2~\cite{resnetse34v22021} system. The EER is 1.03\%.
\end{itemize}
The EER evaluations confirm that the models match their performance expectations.


\noindent \textbf{Over-the-air adversarial attack systems}~We examine the following three systems in the experiments,
\begin{itemize}
    \item \textbf{Baseline:} The baseline is the pipeline in Fig. \ref{fig:pipeline} but without a replay simulator.
    \item \textbf{NRS-only:} The proposed system is presented in Fig.~\ref{fig:pipeline}. When estimating the perturbation, we follow the framework presented in Fig.~\ref{fig:pipeline}. 
    \item \textbf{NRS-joint:} When estimating the perturbation, we follow the framework presented in Fig.~\ref{fig:joint_pipeline}, where both digital attack and NRS are considered in the framework.
    
        Perturbations are synthesized with the PGD attack configuration $\epsilon = 0.0004$ and \textbf{\textit{stop at the first successful attack for a fair comparison}}.  


\end{itemize}

\noindent \textbf{Evaluation metric}~
We evaluate the performance of adversarial examples by the attack success rate metric, and a higher attack success rate indicates more effective attacks and should be of more concern:
\begin{equation}     \label{equ:success_rate}
    \text{Success Rate} = \frac{\text{Number of Successful Attacks}}{\text{Number of Attacks}}
\end{equation}

\subsection{Comparison of NRS loss configurations}
\vspace{-2mm}

We first investigate the effectiveness of different loss configurations when using the NRS to simulate over-the-air attacks. It is achieved by examining the success rates of the white-box over-the-air attack with different NRS loss configurations. 
\vspace{-1mm}
\begin{table}[htbp] 
  \centering
  \caption{Performance of over-the-air white-box attacks in terms of success rates (\%). A higher success rate implies an effective attacking algorithm or loss function.   }
   \resizebox{\linewidth}{!}{
    \begin{tabular}{lcccc|c}
    \toprule
    {NRS Loss} & {XVec} & {RawNet} & TDNN  & ResNet & Average \\
    \midrule
    Baseline & 90.8 & 2.1 & 67.4 & 53.1 & {53.3}\\
    \midrule
    \midrule
    Wav L1  & 88.4 & \textbf{2.4} & 66.5 & 51.6 & {52.2} \\
    Wav L2 & 79.2 & 2.2 & 50.3 & 38.2 & {42.5} \\
    GAN loss & 94.6 & 2.0 & 68.9 & 55.2 & {55.2} \\
    Mel L1 & 93.6 & 2.2 & 71.5 & \textbf{58.1} & {56.4} \\
    \quad w/ Wav L1  & 91.1 & 1.8 & 64.9 & 51.5 & {52.3} \\
    \quad w/ ASV loss & \textbf{96.0} & 1.6 & \textbf{75.3} & 57.8 & \textbf{57.7} \\
    \bottomrule
    \end{tabular}%
    }
    \label{tab:loss_eval_table}
\end{table}%

The results are presented in Table~\ref{tab:loss_eval_table}. The baseline that does not consider over-the-air attacks achieves an average success rate of 53\%. Results implie that attacking the XVec system is the easiest at a success rate of 90.8\%, while it is harder to attack the RawNet system only at a success rate of 2.1\%.
We suspect that because RawNet takes waveform as input, the feature learning is more robust to adversarial attacks. We leave further investigations as a future work.

By integrating the proposed NRS module, the average success rates are increased to 56.4\% and 57.7\% with the Mel L1 loss and Mel L1 + ASV loss, respectively. On the XVec system, GAN loss and Mel L1 + ASV loss can increase success rates from 90.8\% to 94.6\% and 96.0\%, respectively. Mel L1 loss can increase the success rate from 53.1\% to 58.1\% for the ResNet system. However, Wav L1 and Wav L2 losses are only effective to RawNet. We note that RawNet takes raw waveform as input, and Wav L1 and Wav L2 are losses applied directly to waveforms.

In summary, the Mel L1 loss achieves consistent performance in terms of success rates, and considering that the Mel L1 loss 
has been commonly adopted in previous studies on neural vocoders ~\cite{tts_survey2021,hifigan2020}, in the following experiments, we use the neural replay simulator trained on Mel L1 loss for further experiments.

\vspace{-1mm}
\subsection{Performance of the digital and OTA joint attack}
\vspace{-2mm}

\begin{table}[htbp] 
  \centering
  \caption{Performance of the joint attack framework in the context of white-box attack. The success rates (\%) are reported for both digital and over-the-air (OTA) attacks. }
   \resizebox{\linewidth}{!}{
    \begin{tabular}{cccccc|c}
    \toprule
    Attack & Attack & \multirow{2}{*}{XVec} & \multirow{2}{*}{RawNet} & \multirow{2}{*}{TDNN}  & \multirow{2}{*}{ResNet} & \multirow{2}{*}{{Average}} \\
    Path&Framework&~&~&~&~&~ \\
    \midrule
    \multirow{3}{*}{Digital} &Baseline& 100 & 100 & 100 & 100 & 100 \\
    ~ &NRS-only& 100 & 100 & 100 & 100 & 100 \\
    ~ &NRS-joint& 100 & 100 & 100 & 100 & 100 \\
    \midrule
    \multirow{3}{*}{OTA} & Baseline & 90.8 & 2.1 & 67.4 & 53.1 & 53.3\\ 
    &NRS-only & 93.6 & 2.2 & 71.5 & 58.1 & 56.4 \\
    &NRS-joint & \textbf{99.8} & \textbf{4.9} & \textbf{98.0} & \textbf{91.5} & \textbf{73.6} \\
    \bottomrule
    \end{tabular}%
    }
    \label{tab:joint_attack_results}
    \vspace{-2mm}
\end{table}%

We then examine the robustness of the joint adversarial attack framework as illustrated in Fig.~\ref{fig:joint_pipeline}. In this framework, the perturbation is supposed to attack successfully in the context of OTA and digital attacks. Since all frameworks estimate adversarial perturbations until the first success, it is not a surprise that the success rates in digital attacks are all 100\%. However, when performing OTA attacks using the adversarial samples, the baseline only achieves a success rate of 53.3\%. NRS-only and NRS-joint frameworks achieve 56.4\% and 73.6\%, respectively. NRS-joint can increase an absolute success rate of 20.3\%. This implies a 38\% higher chance of attack success compared to the baseline. \textbf{\textit{The experimental results indicate that the proposed joint attack framework can considerably improve the effectiveness of over-the-air adversarial attacks}}.

\vspace{-1mm}
\subsection{Performance of digital and OTA transfer attacks}
\vspace{-3mm}

\begin{table}[htbp]
    \centering
    \caption{Performance of the joint attack framework in the context of transfer attacks. The success rates (\%) are reported for both digital and over-the-air (OTA) attacks.}
   \resizebox{\linewidth}{!}{
    \begin{tabular}{cccccc|c}
    \toprule
    Attack & Attack & \multirow{2}{*}{XVec} & \multirow{2}{*}{RawNet} & \multirow{2}{*}{TDNN}  & \multirow{2}{*}{ResNet} & \multirow{2}{*}{Average} \\
    Path&Framework&~&~&~&~&~ \\
    \midrule
    \multirow{3}{*}{Digital}& Baseline & 44.5 & \textbf{11.5} & 8.9 & \textbf{14.4} & 19.8 \\   
    &NRS-only & 34.4 & 4.4 & 6.8 & 11.3 & 14.2 \\
    &NRS-joint & \textbf{44.7} & {11.3} & \textbf{9.4} & {14.2} & \textbf{19.9} \\    \midrule
    \multirow{3}{*}{OTA}&Baseline & 29.6 & \textbf{4.3} & 5.8 & 9.7 & 12.3\\
    &NRS-only & 28.7 & 3.3 & 5.4 & 8.9 & 11.6 \\
    &NRS-joint & \textbf{32.4} & {3.9} & \textbf{7.3} & \textbf{12.2} & \textbf{14.0} \\   
    \bottomrule
    \end{tabular}%
   }
    \label{tab:physical_transfer}
\end{table}

Last but not least, we examine the performance in the context of transfer attacks. The results are presented in Table~\ref{tab:physical_transfer}. In the context of digital transfer attacks, NRS-only degrades the transfer attack performance, decreasing the average success rate from 19.8\% to 14.2\%, while NRS-joint slightly increases the average success rate from 19.8\% to 19.9\%. 

In the context of OTA transfer attacks, NRS-only slightly decreases the average success rate from 12.3\% to 11.6\%, while NRS-joint increases the average success rate from 12.3\% to 14.0\%. The experimental results indicate that the proposed NRS module and the joint attack framework can increase the success rate for OTA attacks. We need further experiments and analysis to understand why NRS-only decreases the transfer attack performance for both digital and OTA attacks in this transfer attack scenario. 

In summary, \textbf{\textit{the proposed neural replay simulator is able to increase the success rates of OTA adversarial attacks without impacting the success rates of digital adversarial attacks}}. It implies the neural replay simulator is effective in improving adversarial attack robustness. The community should pay more attention to adversarial attacks, and we need to develop more powerful countermeasures.

\vspace{-3mm}
\section{Conclusions} \label{sec:conclusion}
\vspace{-3mm}

In this work, we perform an initial investigation of utilizing neural replay simulators for robust over-the-air adversarial attacks. The experimental results suggest that the proposed neural replay simulator and the joint attack framework can increase the success rate of OTA adversarial attacks. However, there are also limitations of this study. This study only considers one loudspeaker and one microphone. We will leave the generalization study in future work.


\clearpage

\bibliographystyle{IEEEbib_new}
\bibliography{main}

\end{document}